\documentclass[submitting]{nst}
\usepackage{jabbrv}
\usepackage{color}
\usepackage{soul}
\usepackage{subfigure,dcolumn}
\usepackage[T2A,T1]{fontenc}
\usepackage[russian,english]{babel}
\usepackage{listings}
\begin{document}

\title{Characterization and optimization of a cryogenic pure CsI detector with remarkable light yield and unprecedented energy resolution for CLOVERS Experiment}\thanks{This work was supported by the National Key R\&D Program of China 2022YFA1602204, the National Natural Science Foundation of China (Grant Nos. 12175241, 12221005), the Fundamental Research Funds for the Central Universities, the International Partnership Program of the Chinese Academy of Sciences under Grant No. 211134KYSB20200057, and the Double First-Class University Project Foundation of USTC. The authors thank Hefei Comprehensive National Science Center for their strong support.}

\author{Chen-Guang Su}
\affiliation{School of Physical Sciences, University of Chinese Academy of Sciences, Beijing, 100049, China}
\author{Qian Liu}
\email[Corresponding author,]{Qian Liu, liuqian@ucas.ac.cn, +8618910949066}
\affiliation{School of Physical Sciences, University of Chinese Academy of Sciences, Beijing, 100049, China}
\author{Ling-Quan Kong}
\affiliation{School of Physical Sciences, University of Chinese Academy of Sciences, Beijing, 100049, China}
\author{Chen Shi}
\affiliation{School of Physical Sciences, University of Chinese Academy of Sciences, Beijing, 100049, China}
\author{Kimiya Moharrami}
\affiliation{School of Physical Sciences, University of Chinese Academy of Sciences, Beijing, 100049, China}
\author{Yang-Heng Zheng}
\affiliation{School of Physical Sciences, University of Chinese Academy of Sciences, Beijing, 100049, China}
\author{Jin Li}
\affiliation{School of Physical Sciences, University of Chinese Academy of Sciences, Beijing, 100049, China}

\begin{abstract}

In this study, we comprehensively characterized and optimized  a cryogenic pure CsI (pCsI) detector. We utilized a {$\SI{2}{cm}\times\SI{2}{cm}\times\SI{2}{cm}$} cube crystal coupled with a HAMAMATSU R11065 photomultiplier tube,  achieving a remarkable light yield of \SI{35.2}{PE/\keV_{ee}} and an unprecedented energy resolution of \SI{6.9}{\%} at {\SI{59.54}{\keV}}. Additionally, we measured the scintillation decay time of pCsI, which was significantly shorter than that of CsI(Na) at room temperature. Furthermore, we investigated the impact of temperature, surface treatment, and crystal shape on light yield. Notably, the light yield peaked at approximately \SI{20}{\K} and remained stable within the range of \SI{70}--\SI{100}{\K}. The light yield of the polished crystals was approximately 1.5 times greater than that of the ground crystals, whereas the crystal shape exhibited minimal influence on the light yield. These results are crucial for the design of the \SI{10}{\kg} pCsI detector for the future CLOVERS (Coherent eLastic neutrinO(V)-nucleus scattERing at China Spallation Neutron Source (CSNS)) experiment.`
\end{abstract}


\maketitle

\section{Introduction}\label{intro}
The Coherent Elastic Neutrino-Nucleus Scattering (CE$\nu$NS) process has garnered considerable attention since it was first detected in 2017 by the COHERENT collaboration~\cite{akimov2017observation}. Owing to the coherent enhancement of the cross-section (2--3 orders of magnitude higher than any other neutrino matter interaction process with neutrino {energies} below \SI{100}{\MeV}) and nearly pure electromagnetic-weak dynamics (the cross-section is easy to calculate), the CE$\nu$NS process serves as a valuable probe and novel neutrino detection method. As a probe with precise cross- section measurement, it examines the Standard Model at low momentum transfer, aids the understanding of core-collapsed supernova bursts~\cite{ott2013core,tamborra2013neutrino}, helps to determine the neutron radius of a nucleus~\cite{Huang:2024jbh}, and clarifies neutrino fog in WIMP dark matter searches, such as CDEX and PandaX experiments\cite{CDEX:2013kpt,PandaX:2014mem}. As a new neutrino detection method, it offers a flavor-independent approach for searching sterile neutrinos {because it is insensitive to neutrino flavors but sensitive to all flavors of neutrinos~\cite{freedman1974coherent}}, sensitive detection of solar and supernova neutrinos, and threshold-free method to measure the reactor neutrino spectrum below the \SI{1.8}{\MeV} Inverse Beta Decay threshold, which would be highly valuable for reactor oscillation experiments such as JUNO~\cite{JUNO:2015zny}. However, the typical observable ionization energy of a recoiled nucleus is only approximately \SI{1}{\keV} electron equivalent (\SI{}{\keV_{ee}}), making signal detection highly challenging. Various technologies have been proposed for detecting CE$\nu$NS signals. For instance, the RELICS project plans to utilize liquid-xenon detectors~\cite{cai2024relics}, whereas the RECODE experiment opts for HPGe detectors~\cite{yang2024recode}.

To precisely measure the cross-section of the CE$\nu$NS process and advance detector technology based on CE$\nu$NS, we propose the CLOVERS (Coherent eLastic neutrinO(V)-nucleus scattERing at the China Spallation Neutron Source (CSNS)) experiment~\cite{Su:2023klh,HuangMY2016}. Cryogenic pure CsI (pCsI) detectors are adopted because of their large cross section, proportional to the square of the neutron number in the nucleus~\cite{drukier1984principles}, and high light yield, reaching \SI{33.5}{photo-electrons (PE) /\keV_{ee}} coupled to photomultiplier tube (PMT)~\cite{Ding2020uxu} and \SI{43.0}{PE/\keV_{ee}} coupled to Si photomultiplier (SiPM)~\cite{Ding:2022jjm}. To better understand and enhance the performance of the cryogenic pCsI detector, a detailed characterization and optimization was conducted. In this study, the light yield, energy resolution, and scintillation decay time of pCsI crystals at\SI{77}{K} were characterized. Moreover, the influence of temperature, crystal shape, and crystal surface treatment on light yield was investigated. A world-leading energy resolution for scintillator detectors was achieved, and a direction for optimizing future CLOVERS \SI{10}{\kg} pCsI detectors was identified.

\section{Experimental setup}\label{sec:exp_set}

\begin{figure}[!htb]
\includegraphics
  [width=0.9\hsize]
  {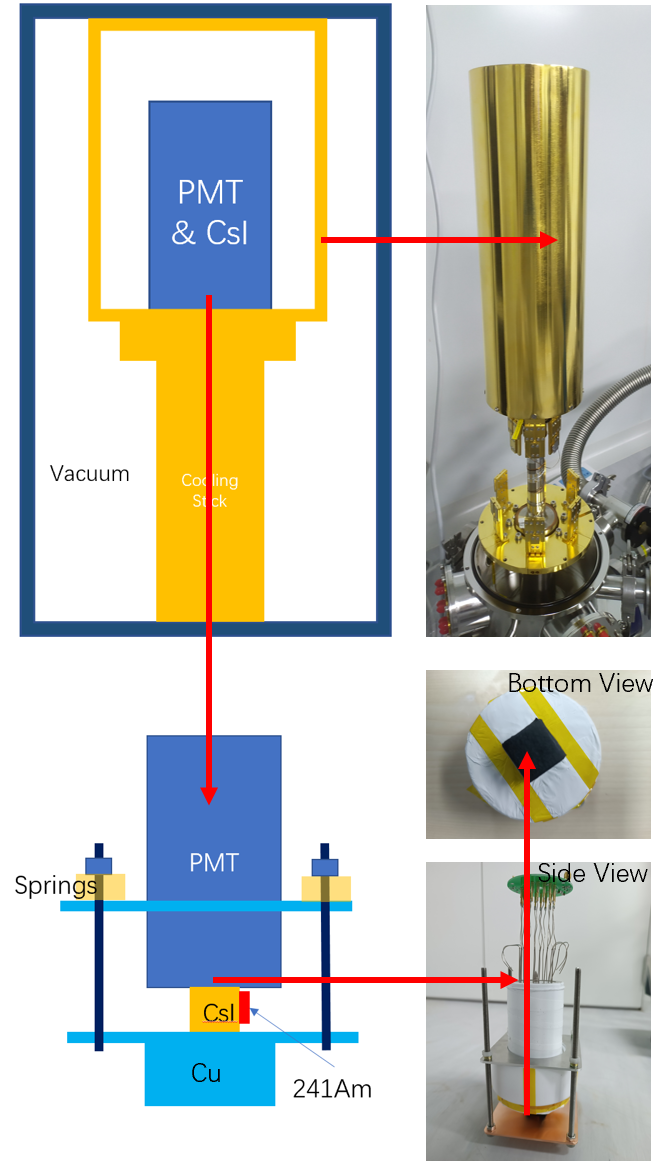}
\caption{(Color online) Sketches and pictures of the experimental setup. A \SI{2}{\cm} cubic pCsI crystal is directly coupled to a HAMAMATSU R11065PMT by springs, {wrapped by four layers of BC-642 Teflon tape and black masking tape from inside out.}}
\label{fig:exp_set}
\end{figure}

The experimental setup used in this study is illustrated in Fig.\ref{fig:exp_set}. The pCsI crystals obtained from HAMAMATSU BEIJING has two different shapes: $\SI{2}{cm}\times\SI{2}{cm}\times\SI{2}{cm}$ cubes and $\Phi2.5\times2$ cm$^{3}$ cylinders. The light output surfaces of all the crystals were polished, whereas the remaining surfaces were ground or polished, depending on the specific crystal. The arithmetic mean roughness values ($\rm R_a$) of the surfaces were approximately \SI{40}{\nm} for polished surfaces and \SI{800}{\nm} for ground surfaces.

An $\rm ^ {241}$Am radioactive source was affixed to the side surface of the crystal. Signals generated by {\SI{10} - \SI{26}{\keV} X-rays and \SI{59.54}{\keV} gamma rays} were used to characterize and optimize the detector. The crystal was directly coupled to the quartz window of a 3-in HAMAMATSU R11065 {because the commonly used optical coupling grease or silicon rubber deforms at low temperatures and deteriorates the light-coupling effect}. To minimize light leakage, the side and bottom surfaces of the crystal, as well as the remaining area of the PMT window, were enveloped in four layers of Luxiumsolutions BC-642 Teflon reflection material. Outside the Teflon-type, the crystal was warped with black masking tape to protect the Teflon layers from stretching. Springs were employed to press the PMT window against the light output surface of the crystal, ensuring adequate optical contact between the PMT and the crystal as well as thermal contact between the crystal and the copper base coupled to the cooling platform using cryogenic vacuum silicone thermal grease outside the black masking tape.

The cryogenic system comprised a custom vacuum chamber, Agilent TPS-mini molecular pump, 
CryoPride KDC6000V helium compressor and CryoPride KDE401SA refrigerating machine capable of achieving a vacuum pressure below \SI{1e-6}{\Pa} and temperature as low as \SI{3}{\K}. Temperature control and monitoring were performed using a Lake Shore Model 325.

The R11065 PMT was powered by a CAEN NDT1470 HV module with a high negative voltage of 1500 V. The PMT signal was read using a CAEN DT5751 digitizer with \SI{1}{\GHz} sampling rate, \SI{500}{\MHz} bandwidth, \SI{1}{Vpp} dynamic range, and \SI{10}{\bit} resolution. The working mode of the DT5751 was set to a self-trigger, and the trigger threshold was set to 30 ADC counts. The original waveform data were directly recorded in the CERN ROOT format using a modified WaveDump software~\cite{CAENOriWaveDump,WaveDumpUCAS}.

\section{Data analysis}\label{sec:algo}

\begin{figure}[!htb]
\includegraphics
  [width=0.9\hsize]
  {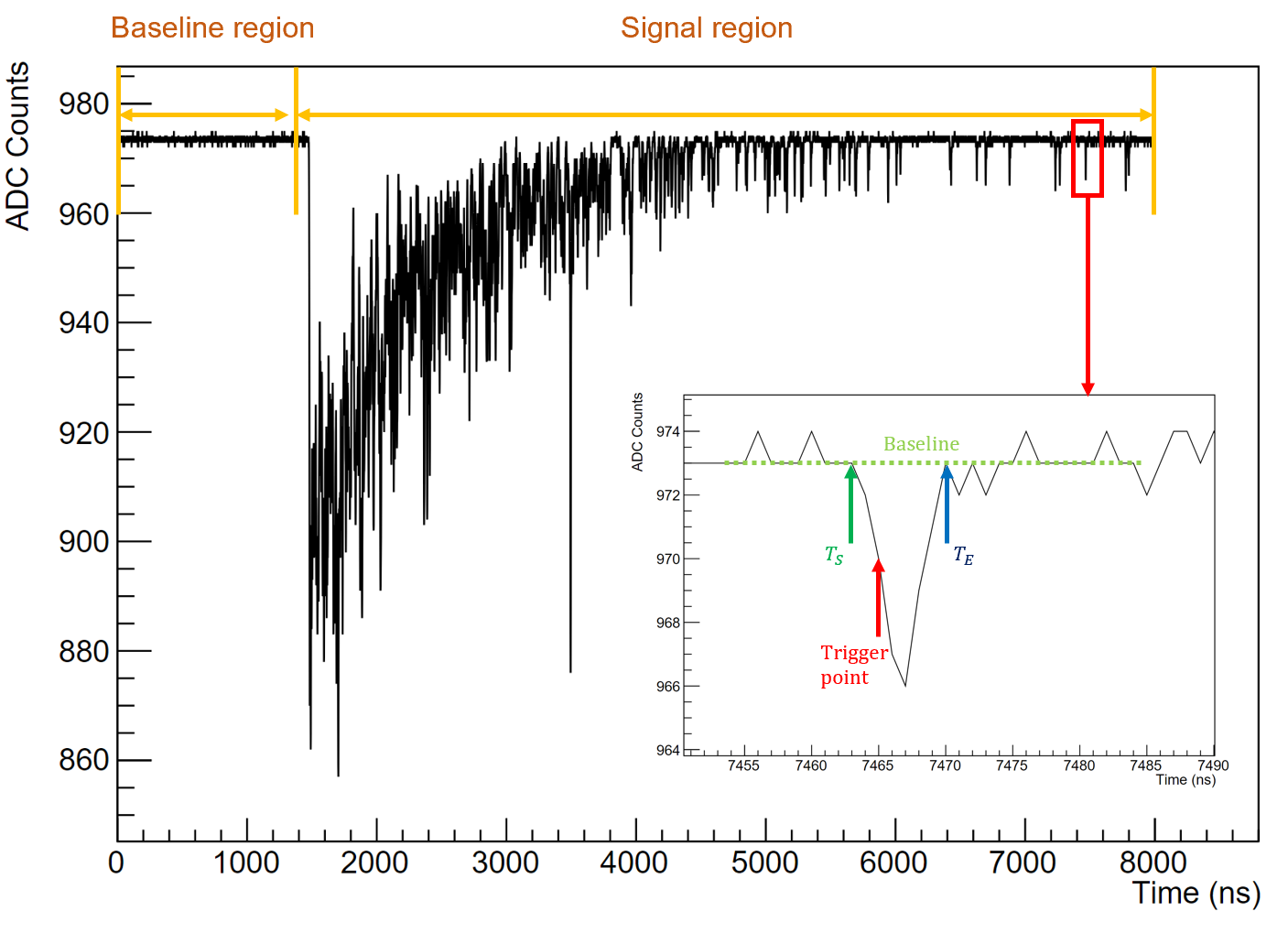}
\caption{(Color online) A typical signal waveform induced by $\rm ^ {241}$Am {\SI{59,54}{\keV}} gamma for pCsI at \SI{77}{\K}. The subplot shows the process of peak searching.}
\label{fig:waveform}
\end{figure}

A typical signal waveform induced by {\SI{59.54}{\keV}} gamma from $\rm ^ {241}$Am in pCsI at \SI{77}{\K} is shown in Fig.\ref{fig:waveform}. Each waveform consists of 8000 samples, corresponding to a duration of \SI{8000}{\ns}, given the \SI{1}{\GHz} sampling rate of DT5751. 

To analyze the waveforms, a toolkit based on C++ and CERN ROOT was developed~\cite{WaveAna} by implementing the following algorithm: 
\begin{enumerate}
    \item The waveform is divided into two regions: the baseline (BL) region covering samples 0--1399 and the signal (SG) region covering samples 1400--7999.
    \item In the BL region, the mean value ($\mu_{\rm BL}$) and standard deviation ($\sigma_{\rm BL}$) of the baseline for each waveform are decided.
    \item Next, the baseline of the waveform is adjusted to 0, and the waveform is inverted.
    \item A peak-searching algorithm to identify all peaks in the SG region is applied. A peak is identified when a sample deviates from 0 by at least 5$\sigma_{\rm BL}$. This sample is defined as the trigger point. The start ($T_{\rm S}$) and end ($T_{\rm E}$) of each peak are defined by the first sample falling back to 0 backwards and forwards from the trigger point, as illustrated in the subplot of Fig.\ref{fig:waveform}.
    \item The integral of the ADC numbers of each peak is recorded in a C++ std::vector named $PeakQ$ for each event. The total charge, that is, the sum of all elements in $PeakQ$, is stored as a C++ float variable named $TotalQ$ for each event. These variables are utilized to derive the results in subsequent analyses.
\end{enumerate}

\section{Characterization of pCsI detector at \SI{77}{\K}}
Using the setup and analysis methods described in Sec.\ref{sec:exp_set} and sec.\ref{sec:algo}, the light yield, energy resolution and scintillation decay time of pCsI detector at \SI{77}{\K} were measured. The crystal used in this characterization was a $\SI{2}{cm}\times\SI{2}{cm}\times\SI{2}{cm}$ cube, with all polished surfaces. 

\subsection{Single PE calibration of the PMT}
\begin{figure}[!htb]
\includegraphics
  [width=0.9\hsize]
  {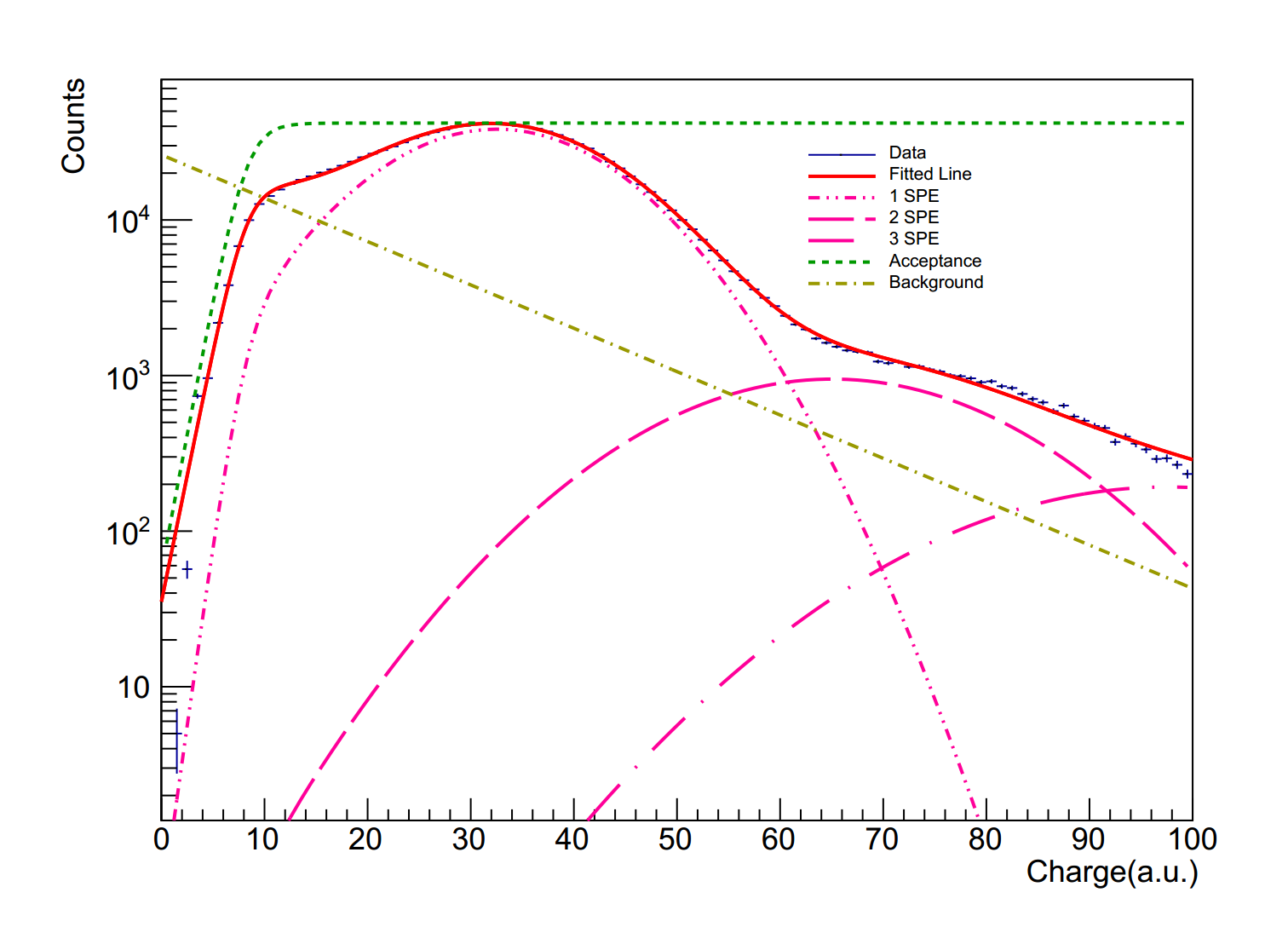}
\caption{(Color online) SPE histogram filled by the PeakQ values of peaks with $T_S$ in range 7000--7999 ns. A simplified SPE model is fitted to the histogram. $\mu_\text{spe} = 32.50\pm0.03$ and $\sigma_\text{spe} = 10.37\pm0.03$ in unit of ADC counts$\cdot$ns}
\label{fig:spe}
\end{figure}

Single-photoelectron (SPE) calibration was necessary to determine the light yield of the pCsI detector. The online SPE calibration was performed by populating a histogram (Fig.\ref{fig:spe}) with the PeakQ values of peaks with $T_S$ between \SI{7000}{\ns} and \SI{7999}{\ns} for all events in dataset. As depicted in Fig.\ref{fig:waveform}, within the $7000-\SI{7999}{\ns}$ range, the pulses are sparse, with each likely corresponding to an SPE. The subplot shows a typical SPE signal. Given the small contributions of multi-PE events, the histogram in Fig.\ref{fig:spe} is fitted using a simplified SPE model, similar to the one described in Ref.\cite{scholz2018first}.
\begin{equation}\label{eq:spefit}
f(q) = [Bkg(q) + \sum_{i=1}^{3}{a_{i}g_{i}(q, Q_\text{spe}, \sigma_\text{spe})] \times Acp(q)} 
\end{equation}
where
\begin{equation}\label{eq:bkg}
Bkg(q) = a_\text{bkg} \cdot e^{-\frac{q}{\sigma_\text{bkg}}}
\end{equation}
\begin{equation}\label{eq:gaus}
g_{i}(q) = {\rm Gaus}(q, iQ_\text{spe}, \sqrt{i}\sigma_\text{spe})
\end{equation}
\begin{equation}\label{eq:acp}
Acp(q) = [1 + e^{-k(q-q_0)}]^{-1}
\end{equation}
The $Bkg(q)$ term describes events generated by electrons emitted from dynodes undergoing incomplete multiplication {as well as some electronic noise}. The $g_i(q)$ terms represent the i-PE Gaussian responses for fully multiplied electrons. $Q_\text{spe}$ is the mean SPE charge and $\sigma_\text{spe}$ represents the spread of SPE charge due to fluctuations in the PMT multiplication process. The $Acp(q)$ term is a sigmoid-shaped function that describes the threshold effect introduced by the peak-searching algorithm discussed in Sec.\ref{sec:algo}. $q_0$ represents the edge midpoint and k controls the steepness of the edge. The $a_i$ values are not constrained to a Poisson distribution, because the process of populating PeakQ in this histogram is not a Poisson procedure.

The fitted curves are shown in Figs.\ref{fig:spe}. The fitted results are as follows: $Q_\text{spe} = 32.50(3)$ and $\sigma_\text{spe} = 10.37(3)$ in units of ADC count $\cdot$ns.

\subsection{Light yield and energy resolution of pCsI}

\begin{figure}[!htb]
\includegraphics
  [width=0.9\hsize]
  {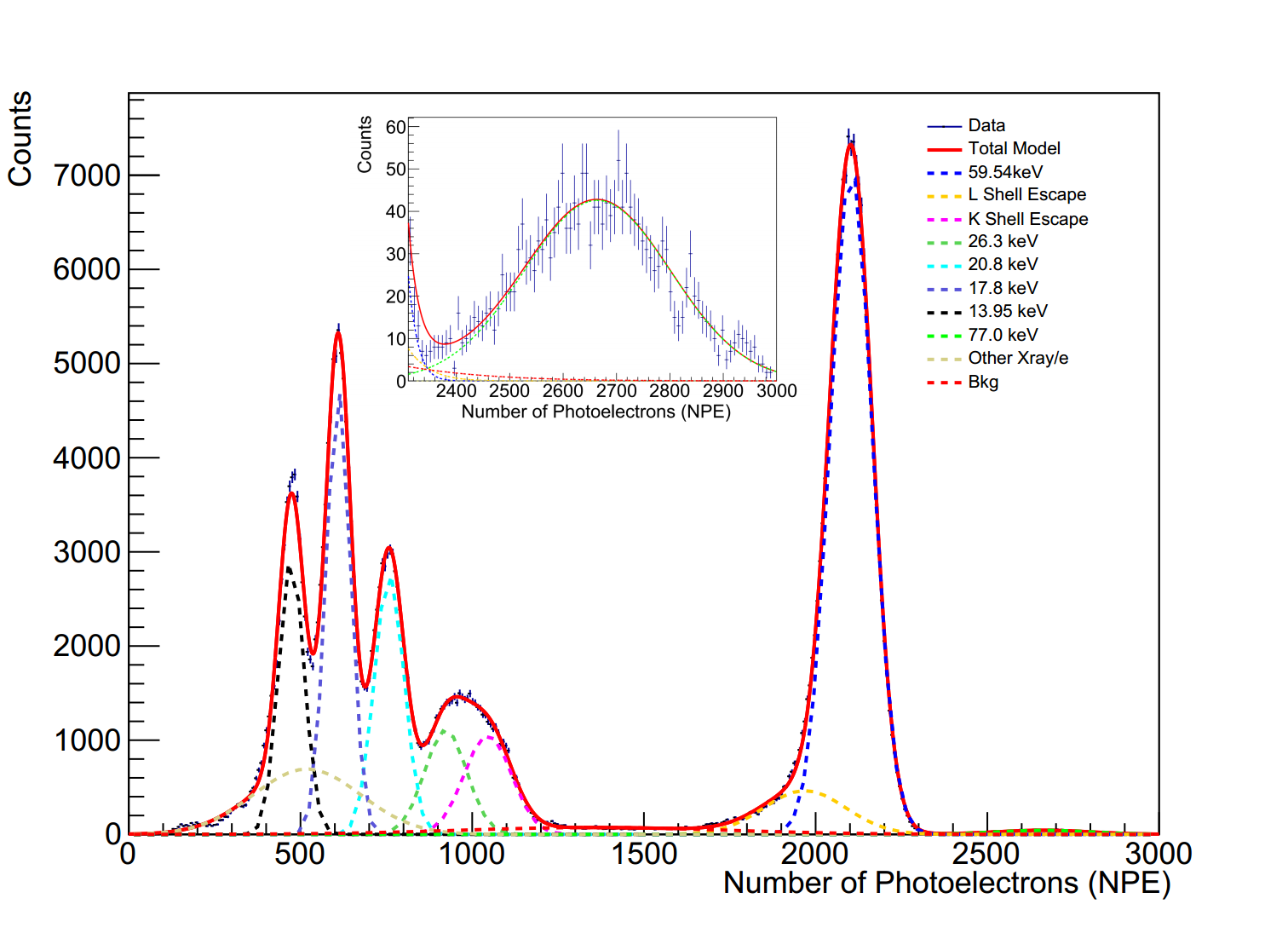}
\caption{(Color online) Energy spectrum of $\rm ^ {241}$Am measured using a pCsI detector at \SI{77}{K}. Ten unconstrained Gaussian functions are used for the global fitting of the spectrum. The fitted curve and its components are shown. {Fitting results shown in Tab.\ref{tab:fit}. The subplot shows the fitting of the \SI{77}{\keV} coincidence peak which is too weak to be visible in the main plot.}}
\label{fig:spec}
\end{figure}

The recorded energy spectrum of $\rm\rm ^ {241}$Am measured in terms of the number of photoelectrons ($NPE$) generated during each event is shown in Fig.\ref{fig:spec} {with the fitting results listed in Tab.\ref{tab:fit}} The $NPE$ is computed using the following formula:
\begin{equation}\label{eq:npe}
NPE = \frac{TotalQ}{Q_\text{spe}}
\end{equation}

The spectrum undergoes a global fit by employing ten unconstrained Gaussian functions to accommodate the various components. The \SI{59.54}{\keV} and \SI{26.3}{\keV} gamma peaks originate from intrinsic gamma rays emitted during $\rm ^ {241}$Am decay. The K-shell escape and L Shell escape peaks are unique to CsI detectors because of the delayed release or escape of X-rays or Auger electrons after a \SI{59.54}{\keV} gamma ejects a K- or L-shell electron into Cs or I atoms. The \SI{13.95}{\keV}, \SI{17.8}{\keV}, and \SI{20.8}{\keV} X-rays originated from the activated states of $^{237}$Np, a decay product of $\rm ^ {241}$Am. These peaks are merged because of nearby X-rays rather than monotonous X-rays.  The \SI{77}{\keV} peak arises from the coincidence of the \SI{59.54}{\keV} gamma-rays and X-rays. The Bkg component accounts for events from an environmentally radioactive background. {
In addition, another X-ray/e component addresses the complex X-ray and electron spectra below \SI{25}{\keV}, requiring more than three Gaussian functions to achieve fitting convergence. However, using a single Gaussian function to describe other X-ray/e spectra remains oversimplified, resulting in an overall $\chi^2/ndf$ value of 4.0. However, when the $\chi^2/ndf$ calculation is constrained to the range NPE$\in$ [600, 3000], the value improves to 1.8. We did not pursue a more refined model for the Other X-ray/e components because their influence on the fitting of the peaks, which are used to determine the light yield, should be minimal because of the significant prominence of the peaks as long as the fitting converges. }

The fitted curves and their components are shown in Figs.\ref{fig:spec}, with fitting results listed in Tab.\ref{tab:fit}. The detector light yield (LY) at various energy points is calculated as:
\begin{equation}\label{eq:ly}
LY [{\rm PE/{keV_{ee}}}] = \frac{\mu_\text{npe}}{Energy {\rm[keV_{ee}]}}
\end{equation}
where $\mu_\text{npe}$ denotes the Gaussian's fitted mean at a given energy point.

\begin{table*}[!htb]
\caption{Fitting results of the measured $\rm ^ {241}$Am spectrum}
\label{tab:fit}
\begin{tabular*}{16cm} {@{\extracolsep{\fill} } cccccc}
\toprule
Type                  & Enerrgy(keV$_\text{ee}$) & $\mu_\text{npe}$ & $\sigma_\text{npe}$ & LY(NPE/keV$_\text{ee}$) & FHWM (\%) \\ \hline
\midrule
\hline
$\gamma$ & 59.54                & 2103.4                     & 61.46                          & 35.3               & 6.9       \\ \hline
L shell excape        & 55.43\textsuperscript{*}                & 1968.4                     & 119.2                         & 35.5               & 14.3      \\ \hline
K shell excape        & 29.5\textsuperscript{*}                 & 1045.6                     & 68.7                          & 35.4               & 15.5      \\ \hline
$\gamma$ & 26.3                 & 922.8                      & 61.2                          & 35.1               & 15.6      \\ \hline
X-ray                 & 20.8\textsuperscript{$\dagger$}                 & 758.3                      & 43.1                          & 36.5               & 13.4      \\ \hline
X-ray                 & 17.8\textsuperscript{$\dagger$}                  & 610.2                      & 35.5                          & 34.3               & 13.7      \\ \hline
X-ray                 & 13.95\textsuperscript{$\dagger$}                & 473.5                      & 36.7                          & 33.9               & 18.2      \\ \hline
Coincidence           & 77.3                 & 2664.0                     & 137.7                         & 34.5               & 12.2      \\ \hline
\bottomrule
\end{tabular*}

\noindent{\footnotesize{\textsuperscript{*} Averaged among Cs and I atoms\cite{radio}.}}
\noindent{\footnotesize{\textsuperscript{$\dagger$} Mean energy of X-rays nearby~\cite{radio}.}}

\end{table*}

\begin{figure}[!htb]
\includegraphics
  [width=0.9\hsize]
  {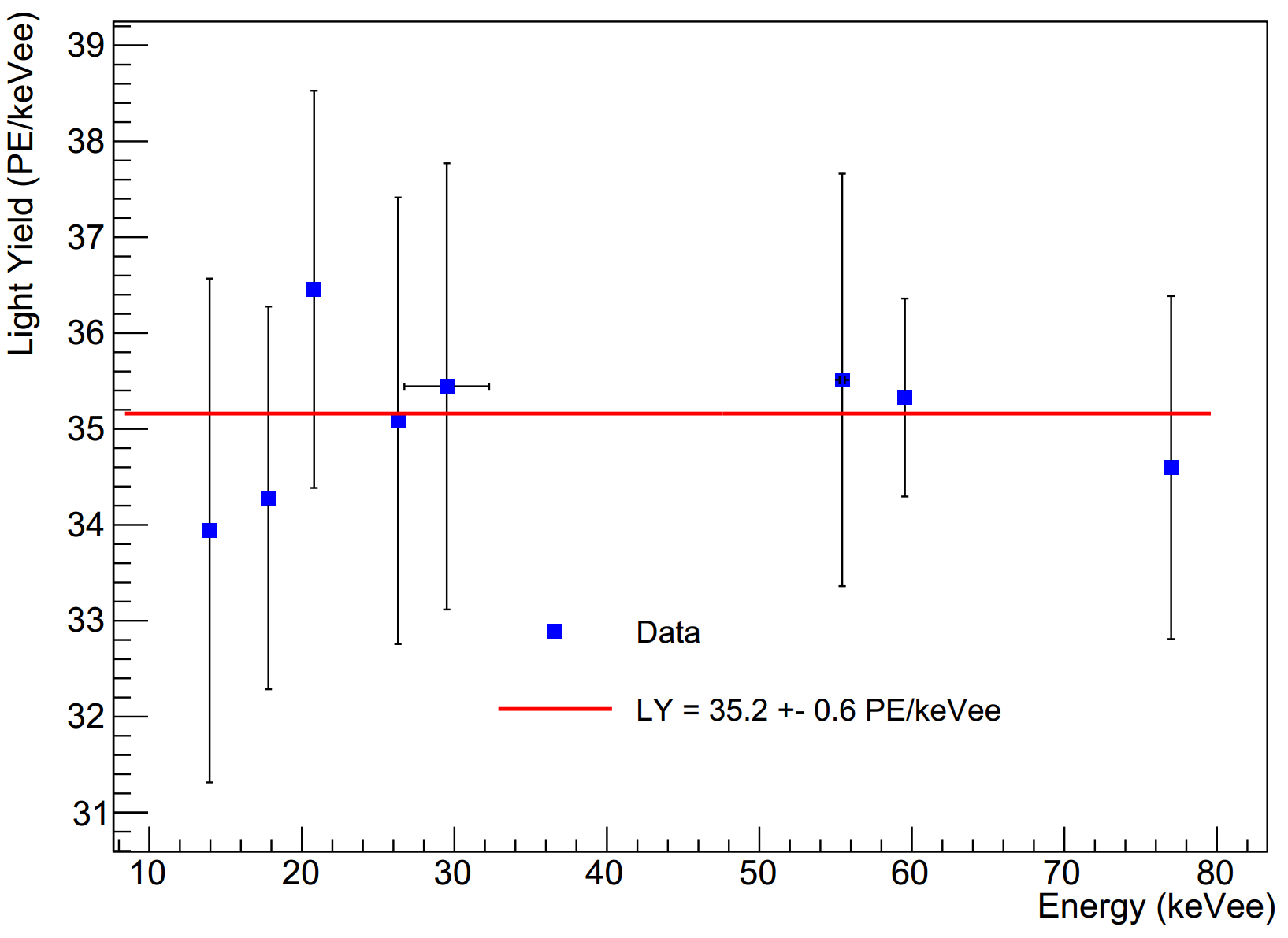}
\caption{(Color online) Light yield calculated at different energy points and their averaged value.}
\label{fig:ly}
\end{figure}

The light yield results obtained at various energy points were fitted using a zero-order polynomial function to determine the mean light yield and the associated uncertainties. The fitted results are shown in Fig.\ref{fig:ly}. {The error bars in the graph represent the standard deviation of the Gaussian function fitted at different energy points divided by the energy ($\sigma_\text{npe}/Energy$).} The final mean light yield of the pCsI detector at \SI{77}{\K} is \SI{35.2\pm0.6}{PE/\keV_{ee}}, which is slightly higher than the assumed value used in the CLOVERS sensitivity estimation, \SI{33.5\pm0.7}{PE/\keV_{ee}}, as reported in Ref.\cite{Ding2020uxu}, where the CsI crystal was also coupled to an R11065 PMT.

\begin{figure}[!htb]
\includegraphics
  [width=0.9\hsize]
  {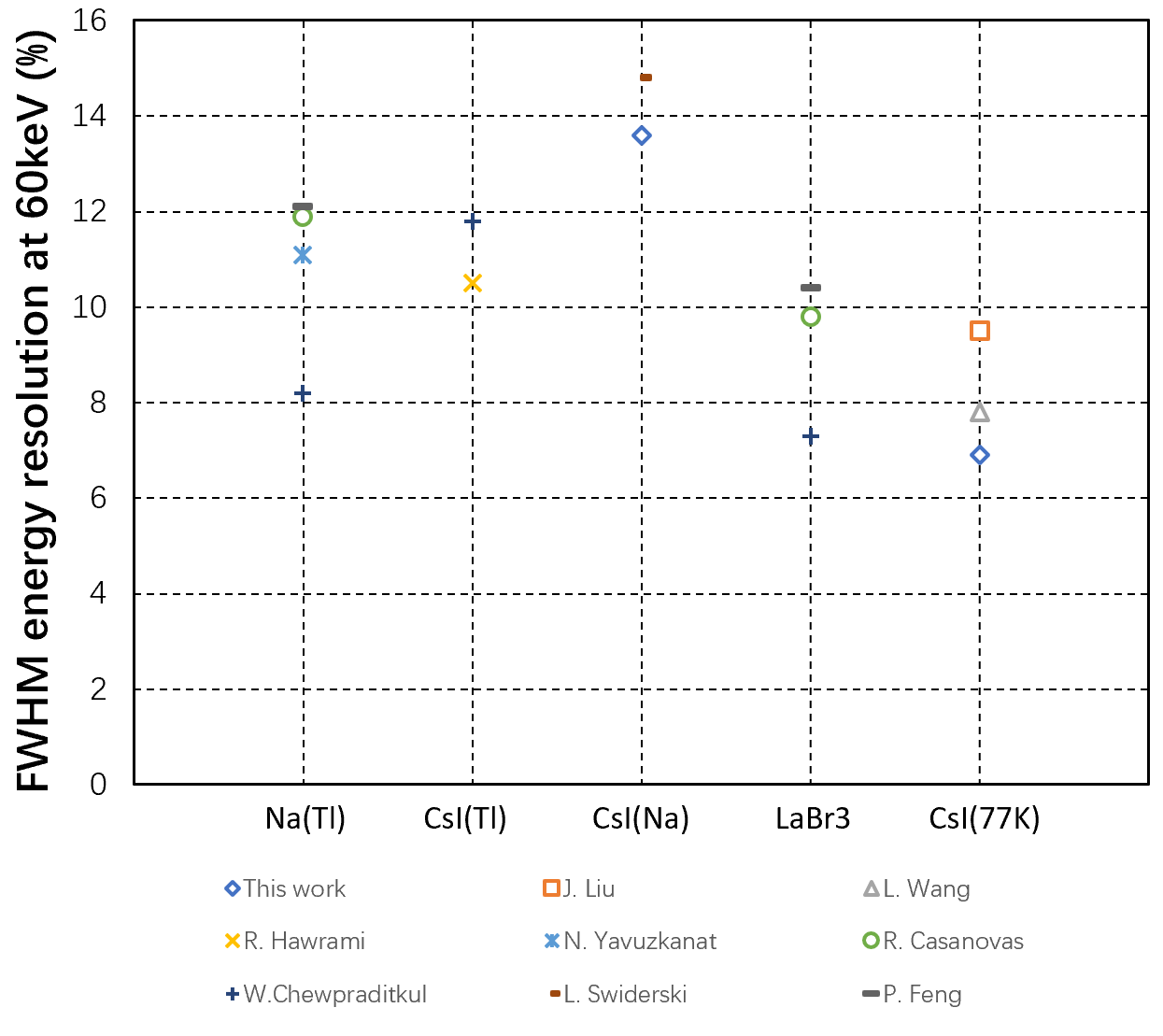}
\caption{(Color online) Comparison of energy resolution at {\SI{59.54}{\keV}} of different crystals from various studies reveals that the \SI{6.9}{\%} resolution achieved in this work by the pCsI detector at \SI{77}{K} is superior to all others. Resolution data were obtained from J. Liu~\cite{Ding2020uxu}, L. Wang~\cite{Wang:2022ekc}, R. Hawrami~\cite{hawrami2022growth}, N. Yavuzkanat~\cite{yavuzkanat2022investigation}, R. Casanovas~\cite{casanovas2012energy}, W. Chewpraditkul~\cite{chewpraditkul2008light}, L. Swiderski~\cite{swiderski2015scintillators} {and P. Feng~\cite{feng2024energy}.}}

\label{fig:res}
\end{figure}

The most remarkable achievement of this study is the world-leading energy resolution achieved by scintillator detectors. Given that the {\SI{59.54}{\keV}} gamma peak is monotonic (unlike the X-rays, K/L Shell escape, and coincidence peaks) and is less influenced by nearby peaks (unlike the \SI{26.3}{\keV} gamma peak, which is strongly affected by nearby X-rays and K shell escape peaks), its resolution is chosen to represent the overall resolution of this study. The full width at half maximum (FWHM) energy resolution of this pCsI detector at \SI{77}{\K} at {\SI{59.54}{\keV}} reached \SI{6.9}{\%}, surpassing the reported \SI{9.5}{\%} in Ref.\cite{Ding2020uxu}, \SI{8.8}{\%} in Ref.~\cite{Ding:2023pqe} and \SI{7.8}{\%} in Refs.\cite{Wang:2022ekc}, making it the best among all the reported resolutions of cryogenic pCsI detectors. This resolution even outperforms that of the brightest inorganic scintillators typically used at room temperature, such as NaI(Tl), CsI(Na), CsI(Tl) or LaBr$_3$. A summary of the comparison is presented in Fig.\ref{fig:res}.

\begin{figure}[!htb]
\includegraphics
  [width=0.95\hsize]
  {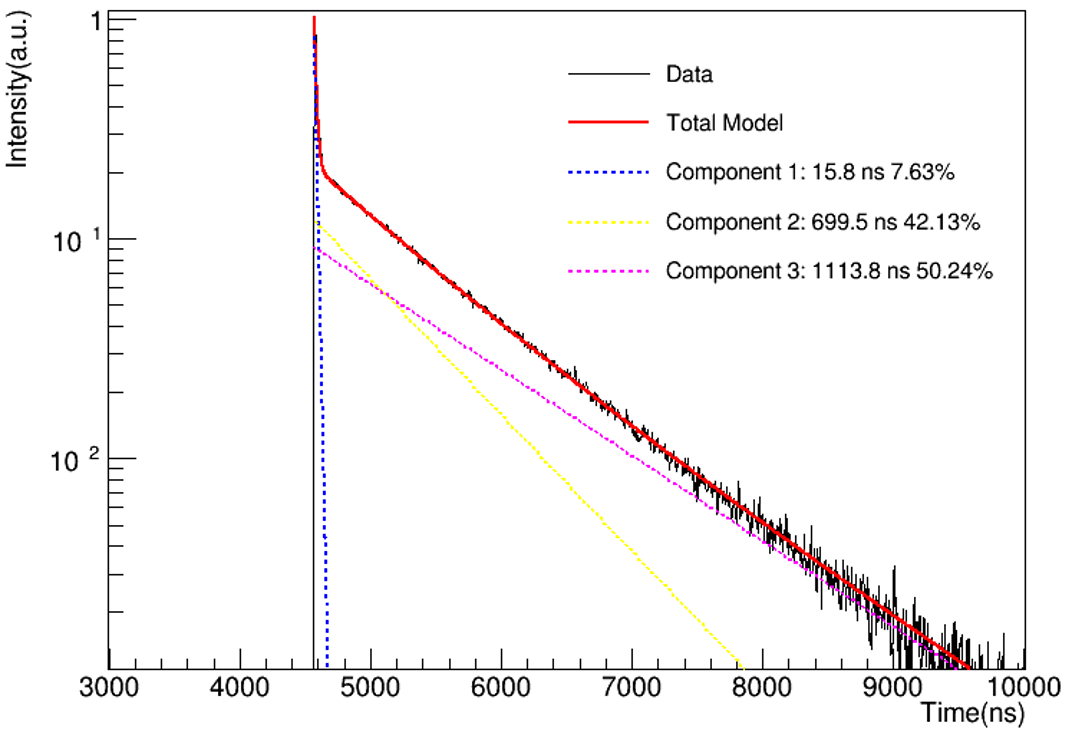}
\caption{(Color online) Scintillation decay time fitting of the pCsI detector at \SI{77}{\K}. The waveform fitted is the sum of 100k $\rm ^ {241}$Am induced events and normalized to its maximum value. The decay constants and composition fractions of different components are listed.}
\label{fig:pCsI_decay}
\end{figure}

\begin{figure}[!htb]
\includegraphics
  [width=0.95\hsize]
  {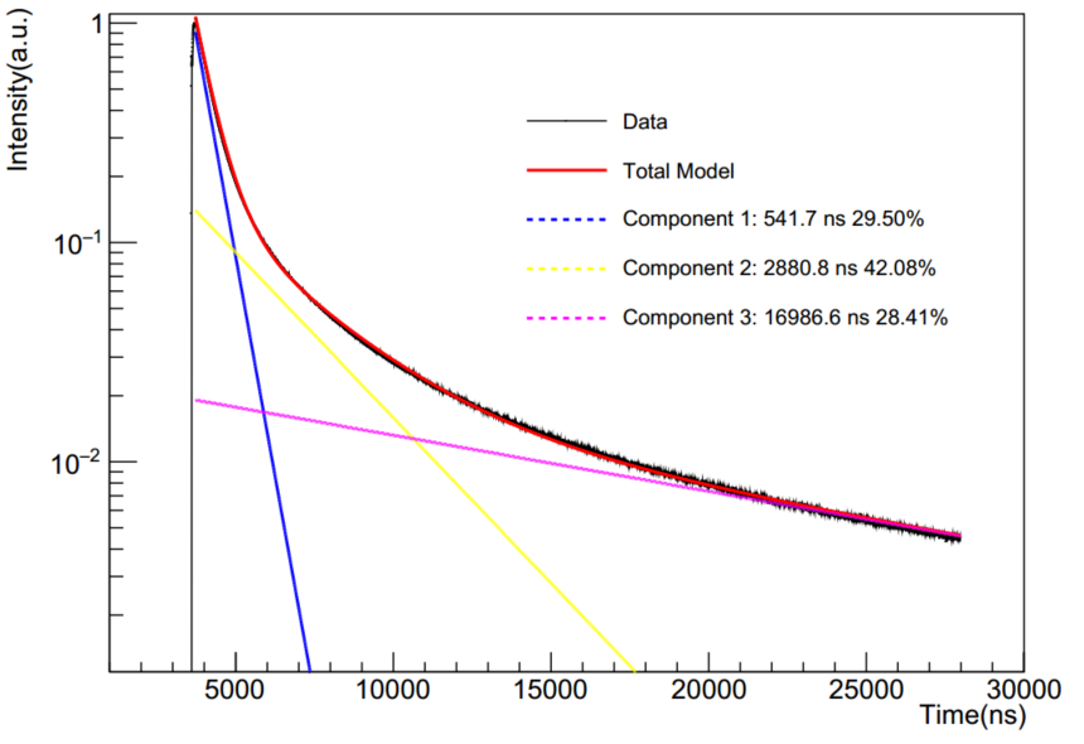}
\caption{Scintillation decay time fitting of the CsI(Na) detector at room temperature followed the same setup as the pCsI experiment, except for the crystal type.}
\label{fig:CsINa_decay}
\end{figure}

\subsection{Scintllation decay time of pCsI}
The scintillation decay time of the pCsI detector at \SI{77}{\K} was determined by fitting the accumulated and normalized waveforms to three exponential components. The waveforms and fitted results are shown in Fig.\ref{fig:pCsI_decay}. The decay time of the slowest component of pCsI at \SI{77}{\K} is approximately \SI{1}{\micro\second}, which is considerably faster than that of the CsI(Na) crystal, which is approximately \SI{17}{\micro\second}, as shown in Fig.\ref{fig:CsINa_decay}. These results are consistent with those in previous studies ~\cite{amsler2002temperature, liu2013luminescence, sun2011fast, Kim:2023lkb}. The considerably shorter decay time of pCsI significantly reduces the afterglow background induced by environmental radioactive events, which contaminates approximately \SI{30}{\%} of all events in the COHERENT CsI(Na) experiment~\cite{scholz2018first}.

\section{Optimization of cryogenic pCsI detector}
To optimize the performance of the cryogenic pCsI detector, investigations were conducted to assess the influence of the temperature, surface treatment, and crystal shape on the light yield.
\subsection{Influence of temperature on the light yield}

\begin{figure}[!htb]
\includegraphics
  [width=0.95\hsize]
  {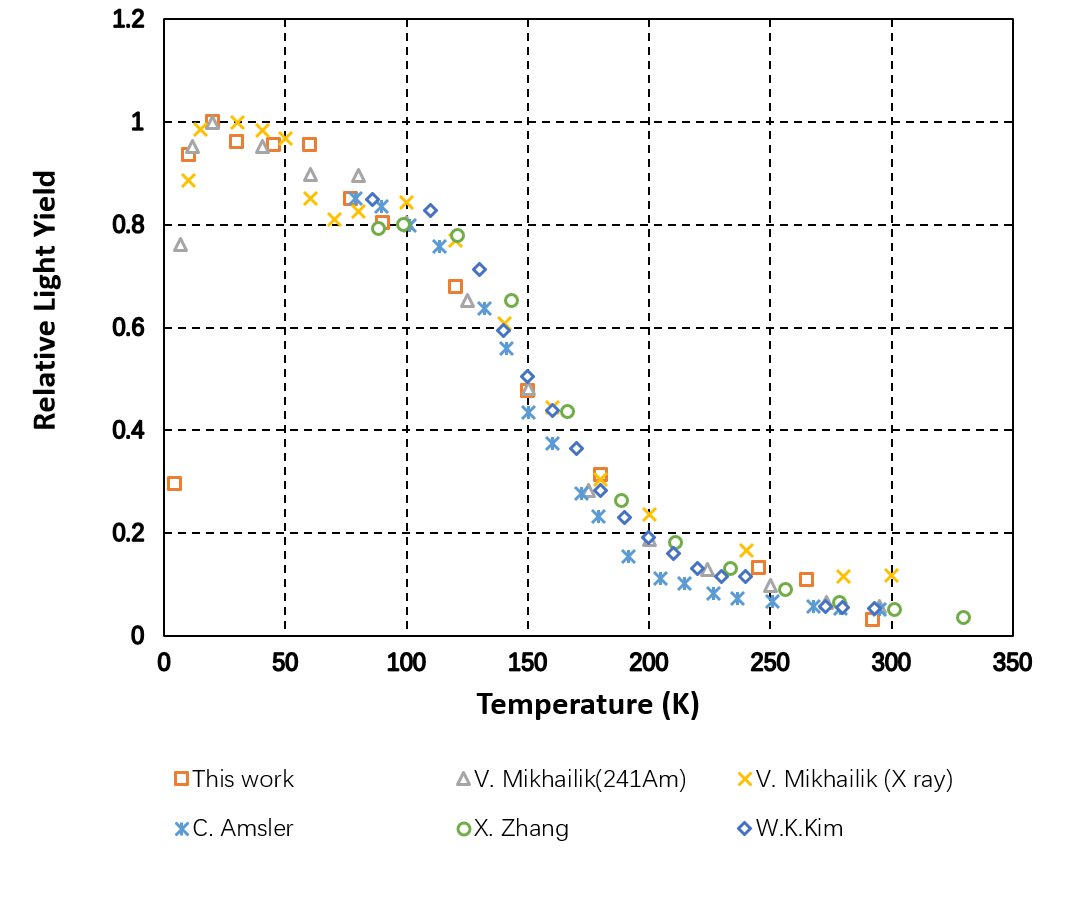}
\caption{(Color online) Relative light yield at various temperatures from multiple datasets shows consistent agreement in the overall trend. Both this study and Mikhailik's results indicate that the maximum light yield is achieved at approximately \SI{20}{\K}. The results from Mikhailik are sourced from Ref.\cite{mikhailik2015luminescence}, Amsler's from Ref.\cite{amsler2002temperature}, Zhang's from Ref.\cite{Zhang2018low} {and W.K.Kim's from Ref.\cite{Kim:2023lkb}.}}
\label{fig:ly_tmp}
\end{figure}

The relative light yields of the pCsI detector at different temperatures are shown in Fig.\ref{fig:ly_tmp}, alongside results from V. Mikhailik~\cite{mikhailik2015luminescence}, C. Amsler~\cite{amsler2002temperature}, X. Zhang~\cite{Zhang2018low} {and W.K.Kim~\cite{Kim:2023lkb}}. The results of this study and those of Mikhailik are normalized to the maximum value for each dataset. However, Amsler and Zhang 
{and W.K. Kim} do not include data points below \SI{77}{\K}; therefore, their light yields are normalized to the values in this study at \SI{77}{\K} based on their results at the same temperature. All results show good agreement in the overall trend, with slight differences, possibly owing to variations in the PMT setup and crystal used.

The light yield increases rapidly as the temperature decreases from room temperature to approximately \SI{100}{\K}. Subsequently, it reached a plateau between \SI{70}{\K} and \SI{100}{\K}, followed by a slight increase at approximately \SI{60}{\K}. The maximum light yield was attained at approximately \SI{20}{\K}, after which it decreased rapidly as the temperature decreased, which is consistent with both our results and those of Mikhailik. Although there is a \SI{20}{\%} increase in light yield from \SI{77}{\K} to \SI{20}{\K}, cooling down to \SI{20}{\K} is considerably more challenging than cooling to \SI{77}{\K}, which can be easily achieved with liquid nitrogen, is inexpensive, and safer than liquid hydrogen, which must be obtained to \SI{20}{\K}.

Because the light yield remains relatively stable between \SI{70}{\K} and \SI{100}{\K}, another potential approach is to place pCsI in an \SI{87}{\K} liquid argon environment and utilize liquid argon as an active veto system to reject the background introduced by external particles such as neutrons or gamma rays.

\subsection{Influence of surface treatment on the light yield}

In addition to temperature, another factor that could influence the light yield is the surface treatment of the crystal, which affects the light-collection efficiency of the detector system, as mentioned in Ref. ~\cite{knyazev2021simulations}]. ~\cite{kilimchuk2010study} and Ref. ~\cite{roncali2013simulation}. To optimize the surface treatment (ground or polished), four comparison experiments were conducted.

\begin{figure}[!htb]
\includegraphics
  [width=0.9\hsize]
  {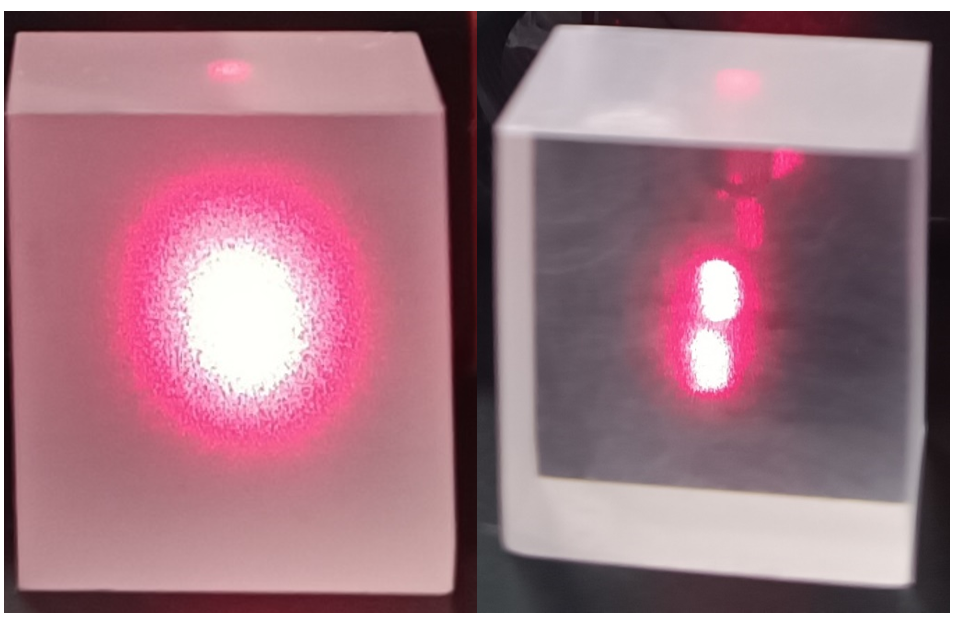}
\caption{(Color online) Laser beam passing through two ground surfaces (left) and two polished surfaces (right)}
\label{fig:surf}
\end{figure}

Experiment A compared the light yields of two cubic crystals subjected to different surface treatments. One crystal had all its surfaces polished, whereas the other had a surface ground, except for the light output surface. The ratio between the light yield of the ground and polished crystals, $R_\text{ly}$, is defined as follows:
\begin{equation}\label{eq:ly_r}
R_\text{ly} = \frac{LY_\text{ground}}{LY_\text{polished}}
\end{equation}
This comparison was conducted at room temperature (\SI{293}{\K}) and \SI{77}{\K}. A laser beam passing through the two ground surfaces and two polished surfaces is shown in Fig. \ref{fig:surf}. The scattering effect of the ground surface on the laser is evident.

Experiment B was identical to Experiment A except that the crystals were cylindrical rather than cubic.

Experiments A and B compared the different crystals. To eliminate differences between crystals, Experiments C and D were conducted. In Experiment C, a cubic crystal was initially ground and then polished, and the light yields were measured before and after polishing. Experiment D was identical to Experiment C, except that a different ground crystal was used. Experiments C and D were conducted at the room temperature. 

In addition to these four experiments with pCsI crystals, similar experiments were conducted for CsI(Na) and CsI(Tl) crystals using the same procedure as in Experiments C and D, respectively. The measured $R_\text{ly}$ from the experiments are listed in Table \ref{tab:r_ly}.

As shown in Table \ref{tab:r_ly}, the light yields of the ground crystals are consistent at approximately \SI{60}-\SI{70}{\%} of the polished crystals, regardless of the temperature, crystal shape, or individual crystals. However, the ratio of CsI(Na) to CsI(Tl) are almost 1. We hypothesize that UV light (\SI{310}{nm} at room temperature ~\cite{pCsI_Spec1,pCsI_Spec2} and \SI{340}{nm} at \SI{77}{K}~\cite{pCsI_Spec1}) emitted by pCsI is more likely to be absorbed by microstructures on the ground surface. In contrast, light with longer wavelengths emitted by CsI(Na) (\SI{420}{nm}~\cite{CsINa_Spec1, CsINa_Spec2}) and CsI(Tl) (\SI{550}{nm}~\cite{CsITl_Spec1, CsITl_Spec2}) is less susceptible to absorption by these microstructures. Although this hypothesis requires further verification, the results suggest that for future \SI{10}{\kg} pCsI detectors in the CLOVERS experiment, all crystal surfaces should be polished to achieve higher light yields.

\begin{table}[]
\begin{tabular}{ccccccc}

\end{tabular}
\end{table}

\begin{table}[!htb]
\caption{Ratio of the light yield ($R_\text{ly}$) between ground and polished crystals for different experiments. Defined in Eq.\ref{eq:ly_r}}
\label{tab:r_ly}
\begin{tabular*}{9cm} {@{\extracolsep{\fill} } ccccccc}
\toprule
Experiment & pCsI(A)    & pCsI(B)    & pCsI(C)    & pCsI(D)   &CsI(Na) &CsI(Tl) \\ \hline
\midrule
\hline
$R_\text{ly}$(293K)    & 0.68 & 0.62 & 0.63 & 0.68 & 0.94 & 1.0\\ \hline
$R_\text{ly}$(77K)     & 0.70 & 0.68 & -    & -   & - & - \\ \hline
\bottomrule
\end{tabular*}
\end{table}

\subsection{Influence of crystal shape on the light yield}
The crystal shape may also influence the light yield, as reported in Ref.\cite{danevich2014impact} and Ref.\cite{sasano2013geometry}. The light yields of cubic and cylindrical crystals were compared for both the ground and polished crystals. The results are presented in Table \ref{tab:ly_shape}. The light yields and energy resolutions of the cubic crystals are slightly higher than those of the cylindrical crystals under both ground and polished conditions; however, the difference is not significant. Therefore, in future CLOVERS experiments, the choice of the crystal shape may depend on the shape of the photon detector to balance the light collection efficiency and detector mass. For instance, the crystal should be cuboidal for SiPMs with square cathode regions and cylindrical for PMTs with circular cathode regions.

\begin{table}[!htb]
\caption{Comparison of the light yield and energy resolution of crystals with different shapes and surface treatment.}
\label{tab:ly_shape}
\begin{tabular*}{8cm} {@{\extracolsep{\fill} } ccccc}
\toprule
Crystal                & LY(PE/keV$_\text{ee}$) & FWHM(\%) \\ \hline
\midrule
\hline
Cubic (polished)         & 35.2               & 6.9      \\ \hline
Cubic (ground)       & 24.8               & 7.8      \\ \hline
Cylindrical (polished)   & 33.9               & 7.1      \\
\hline
Cylindrical (ground) & 22.3               & 7.9       \\
\bottomrule
\end{tabular*}
\end{table}

\section{Summary}

We measured the light yield, energy resolution, and scintillation decay time of pCsI detectors at \SI{77}{\K} coupled with PMT HAMAMASTU R11065. We achieved a light yield of \SI{35.2}{PE/\keV_{ee}}, surpassing the assumed value for CLOVERS sensitivity estimation, and an unprecedented energy resolution of \SI{6.9}{\%} FWHM at {\SI{59.54}{\keV}}, which is the best ever for cryogenic pCsI detectors and world-leading scintillation detectors. The improved light yield and resolution enhanced the sensitivity of CE$v$NS detection. The shorter scintillation decay time of pCsI at \SI{77}{\K} compared with that of CsI(Na) at room temperature {also implies that it is much less likely for afterglow photons from a previous environmental radioactive event to fake a real CE$\nu$NS signal, suppressing this kind of background significantly.}

To optimize the future CLOVERS \SI{10}{\kg} pCsI detector, we investigated the effects of temperature, surface treatment, and crystal shape. Although the light yield peaks at approximately \SI{20}{\K}, cooling a \SI{10}{\kg} material to \SI{20}{\K} was challenging. As the light yield remained stable between \SI{70} and \SI{100}{\K}, placing pCsI in liquid argon at \SI{87}{\K} may be a more viable option, utilizing liquid argon as an active veto system to mitigate the outer background.

Surface treatment significantly affected the light yield, with ground crystals yielding only \SI{60}-\SI{70}{\%} polished crystals. However, the crystal shape had a minimal influence on the light yield. These findings suggest that for future CLOVERS \SI{10}{\kg} pCsI detectors, crystals should be polished, and their shape should match the cathode area of the photon detectors to maximize light yield and utilize the entire sensitive area of the photon detector.

\bibliography{ref.bib}
\bibliographystyle{nst}

\end{document}